\documentclass[runningheads]{llncs}
\usepackage[numbers,sort&compress]{natbib}
\usepackage{graphicx}
\usepackage{amsmath} 
\usepackage{xcolor}
\usepackage{amssymb}

\usepackage{etoolbox}
\patchcmd{\bibliography}{\chapter*}{\section*}{}{}

\begin{document}
\title{Trade-Offs in FMCW Radar-Based Respiration and Heart Rate Variability}
%\author{No Author Given}
\author{Silvia Mura\inst{1} \and
Davide Scazzoli\inst{1} \and
Lorenzo Fineschi\inst{1} \and Maurizio Magarini\inst{1}}
\authorrunning{S. Mura et al.}
%\institute{No Institute Given}
\institute{Dipartimento di Elettronica, Informazione e Bioingegneria,\\ Politecnico di Milano, Milan, Italy, email: name.surname@polimi.it\\
\textbf{Corresponding author}: silvia.mura@polimi.it}
\maketitle          
\begin{abstract}
This study presents a comprehensive experimental assessment of a low-cost frequency-modulated continuous-wave (FMCW) mul-tiple-input multiple-output (MIMO) radar for non-contact vital sign monitoring, focusing on respiratory rate (RR) and heart rate (HR) estimation. The influence of sensing distance and number of transmitted chirps on measurement accuracy is systematically quantified. Results exhibit a U-shaped error profile with optimal performance near $70~cm$, achieving mean absolute errors of $0.8~bpm$ for RR and $3.2~bpm$ for HR. Accuracy deteriorates at short ($<60~cm$) and long ($>100~cm$) distances due to multipath, near-field, and signal-to-noise effects. Increasing chirp count enhances performance: RR errors converge asymptotically for $\geq96$ chirps, while HR requires at least 96 chirps for stable detection. Variability metrics, including heart and respiratory rate variability, remain less accurate ($>15$--$30\%$ error), indicating limited capability in capturing instantaneous fluctuations. These findings define a fundamental trade-off: the radar ensures robust estimation of average RR and HR but exhibits restricted precision in high-resolution beat-to-beat and breath-to-breath monitoring.
\end{abstract}


\section{Introduction}
Continuous monitoring of vital signs, particularly heart rate (HR) and respiratory rate (RR), is crucial for the early detection of physiological anomalies, as subtle fluctuations in these parameters often precede acute clinical events. Timely recognition of such changes substantially improves patient outcomes. Conventional monitoring relies on wearable devices, which provide continuous assessment during daily activities~\cite{singh2020multi}, \cite{dias2018wearable}.

In specific clinical scenarios, wearable sensors are impractical or contraindicated. Patients with contagious diseases risk increased infection or disrupted isolation protocols, while individuals with severe burns or skin injuries experience pain, hindered healing, or contamination from sensors. Patients with psychiatric or cognitive disorders often demonstrate poor compliance, and trauma or emergency care settings limit wearable use due to urgent interventions and equipment constraints. These limitations highlight the need for reliable, contactless vital sign monitoring solutions.

Radar-based systems offer a promising approach for non-contact vital sign (NCVS) monitoring by detecting thoracic micro-movements induced by cardiac and respiratory activity, enabling non-invasive estimation of HR and RR~\cite{tutorial}. Early studies demonstrate the feasibility of extracting respiratory signals using short-range radar, and subsequent research confirms detection of both respiratory and cardiac-induced motion~\cite{gu2009instrument, islam2021can}.

Radar systems for vital sign monitoring fall into continuous-wave (CW) and pulsed categories, including CW Doppler, frequency-modulated continuous-wave (FMCW), step-frequency CW (SFCW), and impulse-radio ultra-wideband (IR-UWB)~\cite{tutorial}. CW Doppler radars are simple and cost-effective but limited in range estimation, whereas wideband techniques such as FMCW, SFCW, and IR-UWB provide more accurate distance measurements. Architectures employ single-input single-output (SISO) or multiple-input multiple-output (MIMO) configurations. SISO systems reduce complexity and cost, while MIMO designs enhance detection accuracy, extend range, and enable multi-target discrimination. Selecting an appropriate radar architecture requires balancing range, accuracy, target separation, complexity, and cost for the intended clinical application\cite{tutorial}.

Millimeter-wave (mmWave) FMCW radar proves particularly effective for NCVS due to higher sensitivity to micro-movements, smaller antennas, and improved target separation~\cite{alizadeh2019remote}. Phased-array implementations steer the beam to detect multiple subjects in distinct angular sectors~\cite{islam2020non}, but angular resolution is limited by array aperture, sequential scanning reduces temporal resolution, and overlapping echoes from closely spaced targets impair accuracy. MIMO radars use multiple transmitter (TX) and receiver (RX) antennas, often with time-division multiplexing, to enhance spatial diversity and isolate targets\cite{su2023time}. Low-cost radars with fewer antennas have reduced spatial resolution and multi-subject discrimination. Hybrid phased-MIMO architectures combine beamforming and MIMO principles to improve angular resolution, target separation, and temporal precision\cite{xu2022simultaneous}.

Despite recent advances in radar-based vital sign monitoring, most existing systems focus primarily on estimating average heart rate (HR) and respiratory rate (RR), offering limited insight into beat-to-beat and breath-to-breath variability. However, variability metrics, such as heart rate variability (HRV) and respiratory rate variability (BRV), are essential clinical indicators of autonomic nervous system activity and cardiopulmonary regulation, providing early markers of stress, fatigue, and pathological conditions\cite{volterrani1994decreased}. Clinical translation of radar-based monitoring remains constrained by motion artifacts, multi-subject interference, phase noise, and environmental clutter. Although techniques such as microprocessor-controlled clutter cancellation~\cite{chen2000microwave}, complex demodulation, motion compensation, adaptive beamforming, harmonic cancellation, wavelet decomposition, and machine learning~\cite{tutorial, zhu2018fundamental, guo2021contactless, xu2022simultaneous} have attempted to mitigate these challenges, the accuracy and reliability of HRV and BRV estimation remain largely unexplored, highlighting a critical gap in understanding the trade-offs between average-rate accuracy and instantaneous physiological variability.

This study addresses this gap by performing a comprehensive experimental evaluation of a low-cost FMCW MIMO radar for non-contact monitoring of RR and HR, explicitly including variability metrics. Our results reveal a clear operational trade-off: RR can be reliably estimated across a broad range of distances ($50-120~cm$) and chirp counts (96--256), whereas accurate HR detection requires optimal positioning near $70~cm$ and at least 96 chirps. Variability metrics, however, remain less precise, with errors typically exceeding 15--30\%, underscoring the current limitations of radar systems in resolving instantaneous physiological fluctuations. These findings suggest that while FMCW MIMO radar is well-suited for robust estimation of average vital signs, achieving clinical-grade resolution of HRV and BRV demands further refinement. Potential improvements may include multi-domain data fusion or temporal super-resolution techniques to improve beat-to-beat sensitivity. By systematically quantifying these dependencies and trade-offs, this work provides actionable guidelines for radar configuration, establishes quantitative performance benchmarks, and extends the literature by explicitly addressing variability metrics as a key factor in non-contact vital sign monitoring.

\section{System model}
This section presents a rigorous mathematical formulation of the MIMO FMCW radar system for contactless vital sign monitoring. We develop a comprehensive model that integrates electromagnetic wave propagation, human physiological dynamics, and signal processing theory to enable precise extraction of cardiorespiratory parameters.

\subsection{Radar Model}
We consider a MIMO FMCW radar system characterized by $N_{TX}$ transmitting antennas and $N_{Rx}$ receiving antennas arranged in a virtual array configuration. The transmitted signal from the 
$n$th TX antenna during the $c$th chirp is modeled as a linear frequency-modulated (LFM) waveform
\begin{equation}
s_{n}(t) = A_{\text{TX}} \exp\left\{ j 2 \pi \left(f_{c} t + \frac{K}{2} t^{2} \right) \right\}, \quad n \in {1, 2, \ldots, N_{\text{TX}}},
\label{eq:transmitted_signal}
\end{equation}
where $f_c$ denotes the carrier frequency, $K$ represents the chirp rate (frequency sweep rate), and $A_{TX}$ is the complex transmitted coefficient.

The human subject is modeled as an extended scatterer located at range $R(t)$ and azimuth angle $\theta(t)$ $\in [-\pi/3, \pi/3]$ relative to the radar array center. The time-varying nature of these parameters captures the micro-motion induced by cardiorespiratory activity. The received signal at the $m$th RX antenna from the $n$th TX antenna is described by
\begin{equation}
r_{mn}(t) = \alpha_{mn}(t) \cdot s_{n}(t - \tau_{mn}(t)) + w_{mn}(t)
\label{eq:received_signal}
\end{equation}
where $\tau_{mn}(t) = \frac{2 R_{mn}(t)}{c}$, $c$ denotes the speed of light, and  
$w_{mn}(t)$ $\sim$ $\mathcal{N}(0, \sigma_{w}^2$) represents circularly symmetric complex additive white Gaussian noise with zero mean and variance $\sigma_{w}^2$. In this model, we assume a propagation environment with no external interference or multipath reflections, such that only the direct target echo contributes to the received signal.

The complex attenuation coefficient $\alpha_{mn}$(t) $\in \mathbb{C}$ encapsulates the comprehensive channel response

\begin{equation}
\alpha_{mn}(t) = \sqrt{\frac{\lambda^{2} \sigma_{\text{RCS}}}{(4\pi)^{3} R_{mn}^{4}(t)}} \exp\left\{-j \frac{4\pi}{\lambda} \Delta d(t)\right\}
\label{eq:attenuation_coeff}
\end{equation}
where $\lambda=c/f_c$ is the wavelength, $\sigma_{\text{RCS}}$ is the radar cross-section of the thoracic region, and $\Delta d(t)$ represents the chest displacement from the baseline position.

Radar measurements are systematically collected and organized into a four-dimensional complex tensor $\mathbf{R} \in \mathbb{C}^{F \times N_{\text{RX}} \times N_{c} \times N_{a}}$, 
where $F$ is the number of temporal frames, $N_c$ is the number of chirps per frame, and $N_a$
is the number of fast-time samples per chirp. The temporal resolution is governed by the frame repetition interval $T_f$, which determines the slow-time sampling frequency $f_{s} = \frac{1}{T_f}$.

\subsection{Cardiovascular and Respiratory Physiology}
The human cardiovascular system comprises two interconnected circuits operating in synchronized cardiac cycles. Each cycle involves coordinated contractions (systole) and relaxations (diastole) producing mechanical vibrations detectable by radar systems.
The respiratory system facilitates gas exchange through cyclical ventilation with typical parameter ranges: \text{HR} $\in$ [60, 100] \text{ beats per minute} 
\text{RR} $\in$ [12, 25] \text{ breaths per minute}, $\delta_{h}$ $\in$ [1, 9] \text{ mm (cardiac displacement)\cite{shafiq2014surface}} and 
$\delta_{b}$ $\in$ [10, 50] \text{ mm (respiratory displacement)}\cite{kaneko2012breathing}. 
The instantaneous chest displacement is modeled as:
\begin{equation}
\Delta d(t) = d(t) - d_{0} = \delta_{b}(t) + \delta_{h}(t) + \epsilon(t)
\label{eq:chest_displacement}
\end{equation}
where $d_0$ is the baseline distance, $\delta_{b}(t)$ and $\delta_{h}(t)$ are respiratory and cardiac displacements respectively, and $ \epsilon(t)$ represents measurement noise and artifacts.
\section{Radar Signal Processing}
This section delineates the systematic signal processing pipeline that transforms raw MIMO FMCW radar measurements into robust estimates of respiratory and heart rates. The following framework, adapted from \cite{tutorial}, has been used
\begin{itemize}
\item\textbf{Two-Dimensional Spectral Analysis}:
for each temporal frame $k$ and chirp $j$, we construct a range--azimuth map through two-dimensional Fourier analysis:
\begin{equation}
\mathbf{A}_{i,j}(r,a) = \text{FFT}_{\text{azimuth}}\!\left\{ \text{FFT}_{\text{range}}\!\left( \mathbf{R}_{k,:,j,:} \right) \right\}.
\label{eq:2d_fft}
\end{equation}
This transformation yields a spatially resolved representation of backscattered energy, enabling precise localization of the thoracic region.

\item\textbf{Optimal Bin Selection Algorithm}:
The selection of the optimal range-azimuth bin $(r^{*}, a^{*})$ is crucial for maximizing sensitivity to physiological micro-motions while minimizing noise and clutter interference. This process involves three systematic steps:
\begin{itemize}
\item Step 1: 
We first restrict the search to physically meaningful locations by defining a feasible region $\Omega$:
\begin{equation}
\Omega = \left\{ (r,a) : r_{\min} \leq r \leq r_{\max}, \quad |\mathbf{A}_{k,j}(r,a)|^{2} \geq  P_{\text{threshold}} \right\}
\label{eq:feasible_region}
\end{equation}
where $r_{\min}$ and $r_{\max}$ define the valid range window (excluding near-field coupling and far-field low-SNR region,$P_{\text{threshold}}$ is a power threshold to exclude low-power static clutter regions.

\item {Step 2: }
For each candidate bin $(r,a) \in \Omega$, we compute two complementary metrics:

\textit{Phase Variability Metric} (measures temporal fluctuations):
\begin{equation}
\sigma_{\phi}(r,a) = \sqrt{\frac{1}{N_{c}-1} \sum_{j=1}^{N_{c}} \left[\phi_{j}(r,a) - \bar{\phi}(r,a)\right]^{2}}
\label{eq:phase_variability}
\end{equation}
where $\phi_{j}(r,a) = \arg(\mathbf{A}_{k,j}(r,a))$ is the phase for chirp $j$, and $\bar{\phi}(r,a) = \frac{1}{N_{c}}\sum_{j=1}^{N_{c}} \phi_{j}(r,a)$ is the mean phase.

\textit{Power Concentration Metric} (measures signal strength):
\begin{equation}
P_{\text{log}}(r,a) = \log\left(\frac{1}{N_{c}} \sum_{c=1}^{N_{c}} |\mathbf{A}_{k,j}(r,a)|^{2}\right)
\label{eq:power_metric}
\end{equation}

\item{Step 3:}
The optimal bin maximizes a weighted combination of both metrics:
\begin{equation}
(r^{*}, a^{*}) = \arg\max_{(r,a) \in \Omega} \left[ w_{1} \sigma_{\phi}(r,a) + w_{2} P_{\text{log}}(r,a) \right]
\label{eq:optimal_bin}
\end{equation}
with weights $w_{1}, w_{2} \geq 0$ and $w_{1} + w_{2} = 1$. Typically, $w_{1} = 0.7$ and $w_{2} = 0.3$ to prioritize phase variability (which indicates motion) over raw power.

The rationale is that chest micro-motions cause significant phase variations across chirps, while static objects exhibit stable phases.
\end{itemize}

\item \textbf{Phase Extraction and Preprocessing}:
Once the optimal bin $(r^{*}, a^{*})$ is identified, we extract the slow-time phase sequence by averaging the complex signal across all chirps within each frame:
\begin{align}
\bar{z}_{k} &= \frac{1}{N_{c}} \sum_{j=1}^{N_{c}} \mathbf{A}_{k,j}(r^{*}, a^{*}) \label{eq:phase_avg} \\
\phi_{k} &= \text{unwrap}\big(\arg(\bar{z}_{k})\big) \label{eq:phase_unwrap}
\end{align}
The averaging operation in (\ref{eq:phase_avg}) improves the signal-to-noise ratio by reducing uncorrelated noise across chirps. The unwrapping operation in (\ref{eq:phase_unwrap}) removes $2\pi$ discontinuities to obtain a continuous phase trajectory that represents the chest displacement over time.

\item \textbf{Respiratory Rate Estimation}:
Respiratory activity manifests as large-amplitude, low-frequency oscillations in the unwrapped phase signal. The estimation procedure follows:
\begin{align}
\phi_{\text{resp},k} &= \mathcal{H}_{\text{BP}}^{(0.1, 0.5)} \left\{ {\phi}_{k} \right\} \label{eq:resp_filter} \\
e_{\text{resp},k} &= \left| \phi_{\text{resp},k} + j \mathcal{H}\left\{ \phi_{\text{resp},k}[ \right\} \right| \label{eq:resp_envelope} \\
\widehat{\text{RR}} &= 60 \left\langle \frac{1}{\Delta t_{\text{peaks}}} \right\rangle \text{ bpm} \label{eq:rr_estimate}
\end{align}
where $\mathcal{H}_{\text{BP}}^{(0.1, 0.5)}$ represents band-pass filtering in the respiratory frequency band (0.1--0.5 Hz, corresponding to 6--30 breaths per minute), $\mathcal{H}\{\cdot\}$ is the Hilbert transform for envelope extraction\cite{linschmann2022estimation}, and $\langle \cdot \rangle$ denotes temporal averaging over detected inter-peak intervals $\Delta t_{\text{peaks}}$.

\item \textbf{Heart Rate Estimation}:
Cardiac signals have much smaller amplitude than respiratory signals and may be contaminated by respiratory harmonics. Therefore, additional preprocessing is required:
\begin{align}
\phi_{\text{notch},k} &= \mathcal{H}_{\text{notch}}^{(f_{\text{BR}})} \left\{ {\phi}_{k} \right\} \label{eq:notch_filter} \\
\phi_{\text{cardiac},k} &= \mathcal{H}_{\text{BP}}^{(0.9, 3.0)} \left\{ \phi_{\text{notch},k} \right\} \label{eq:cardiac_filter}
\end{align}
where $\mathcal{H}_{\text{notch}}^{(f_{\text{BR}})}$ is a notch filter that attenuates the fundamental respiratory frequency $f_{\text{BR}}$ and its harmonics to reduce respiratory interference.

The heart rate is then estimated in the frequency domain using the discrete Fourier transform as in \cite{tutorial}
\begin{align}
X[n] &= \left|\sum_{k=0}^{N_{\text{FFT}}-1} \phi_{\text{cardiac},k}\,e^{-j 2\pi nk/N_{\text{FFT}}}\right| \label{eq:dft} \\
f_{n} &= \frac{n f_{s}}{N_{\text{FFT}}} \label{eq:freq_bins} \\
n^{*} &= \arg\max_{n: f_{n} \in [0.9, 3.0]} X[n] \label{eq:peak_detection} \\
\widehat{\text{HR}} &= 60 f_{n^{*}} \text{ bpm} \label{eq:hr_estimate}
\end{align}
where the frequency search is restricted to the cardiac band [0.9, 3.0] Hz (54--180 bpm) to avoid spurious detections.

It is important to remark that respiration and heartbeat exhibit different characteristics, motivating distinct processing methods. Respiration is a low-frequency, quasi-periodic signal with large amplitude; applying the Hilbert transform yields the analytic signal, from which instantaneous phase and frequency can be extracted for accurate breathing rate tracking\cite{linschmann2022estimation}. Heartbeat, being higher-frequency and lower-amplitude, is masked by respiration; after removing the fundamental respiration frequency and its harmonics via a notch filter, Fourier analysis efficiently identifies the heart rate from spectral peaks. Hence, Hilbert transform suits slow, quasi-periodic respiration, while filtered Fourier analysis is better for fast, periodic heartbeats.

\end{itemize}

\section{Cardiovascular Metrics}
This section presents the metrics analyzed for NCVS monitoring, focusing on both HR and RR. In particular, the HR-related metrics characterize the instantaneous fluctuations within the cardiac cycle, encompassing not only the average values but also statistical descriptors such as the standard deviation. These metrics play a pivotal role in detecting and defining clinical anomalies in cardiac dynamics, providing more comprehensive diagnostic information than the sole estimation of average HR over the measurement period\cite{volterrani1994decreased}. Analogous variability metrics are also evaluated for the respiratory signal to capture fluctuations in the breathing cycle.

\subsection{Heart Rate Variability (HRV) Metrics}
Although the estimation of the heart rate provides very important information, it represents only a limited view of cardiovascular dynamics. A richer characterization is obtained by analyzing the fluctuations in the duration of successive cardiac cycles, a quantity known as Heart Rate Variability (HRV). HRV reflects the interplay between the sympathetic and parasympathetic branches of the autonomic nervous system, and its analysis has become a cornerstone in cardiology, stress research, and psychophysiology.

From the cardiac-filtered signal, Inter-Beat Intervals (IBIs) are extracted and processed to compute standard HRV metrics\cite{shaffer2017overview}:
\begin{align}
\overline{\text{IBI}} &= \frac{1}{N}\sum_{i=1}^{N} \text{IBI}_{i}, \label{eq:mean_ibi} \\
\text{SDNN} &= \sqrt{\frac{1}{N-1}\sum_{i=1}^{N}(\text{IBI}_{i} - \overline{\text{IBI}})^{2}}, \label{eq:sdnn} \\
\text{RMSSD} &= \sqrt{\frac{1}{N-1}\sum_{i=1}^{N-1}(\text{IBI}_{i+1} - \text{IBI}_{i})^{2}}, \label{eq:rmssd} \\
\text{pNN50} &= \frac{100}{N-1} \sum_{i=1}^{N-1} \mathbf{1}_{\{|\text{IBI}_{i+1}-\text{IBI}_{i}| > 50 \ \text{ms}\}}, \label{eq:pnn50}
\end{align}
where $\mathbf{1}_{\{\cdot\}}$ is the indicator function. These metrics are defined as:
\begin{itemize}
    \item \textbf{Mean IBI ($\overline{\text{IBI}}$):} the average cardiac period, directly linked to mean heart rate.
    \item \textbf{SDNN (Standard Deviation of NN intervals):} a global measure of overall variability.
    \item \textbf{RMSSD (Root Mean Square of Successive Differences):} sensitive to parasympathetic activity.
    \item \textbf{pNN50:} the percentage of successive IBIs differing by more than 50 ms, often adopted in clinical practice as an indicator of vagal tone.
\end{itemize}

\subsection{Breath Rate Variability (BRV) Metrics}
A parallel analysis can be carried out for respiratory activity, giving rise to Breath Rate Variability (BRV). Although less standardized than HRV, BRV has recently attracted attention as it reflects not only the mechanical dynamics of breathing but also its modulation by cortical and autonomic mechanisms. From the respiratory envelope, we detect consecutive inspiratory peaks and compute the Breath-to-Breath Intervals (BBIs). After discarding implausible values (outside the 1.5--10 s range) and rejecting outliers, the following indices are extracted as in \cite{soni2019breath}
\begin{align}
\text{MIBI} &= \frac{1}{M}\sum_{j=1}^{M} \text{BBI}_{j}, \label{eq:mibi} \\
\text{SDBB} &= \sqrt{\frac{1}{M-1}\sum_{j=1}^{M}(\text{BBI}_{j} - \text{MIBI})^{2}}, \label{eq:sdbb} \\
\text{RMSSD}_{\text{BBI}} &= \sqrt{\frac{1}{M-1}\sum_{j=1}^{M-1}(\text{BBI}_{j+1} - \text{BBI}_{j})^{2}}, \label{eq:rmssd_bbi}
\end{align}
where
\begin{itemize}
    \item \textbf{MIBI (Mean Inter-Breath Interval):} the average breathing period.
    \item \textbf{SDBB (Standard Deviation of BBIs):} a general descriptor of the dispersion of respiratory cycles.
    \item \textbf{RMSSD$_{\text{BBI}}$:} the short-term variability of breathing cycles.
\end{itemize}

These metrics provide a concise description of respiratory variability, complementing the average breathing rate. High BRV is often associated with resilience and healthy autonomic regulation, while reduced BRV may indicate stress, fatigue, or compromised respiratory control. Although clinical guidelines for BRV are not as standardized as those for HRV, including these indices enriches the overall physiological assessment from radar measurements.

Together, these metrics enable not only the comparison between radar-derived signals and ground truth measurements, but also the evaluation of the radar's ability to capture subtle physiological variability relevant to stress assessment, sleep analysis, and cardiovascular risk stratification.

\section{Experimental Campaign}
This section presents the experimental campaign designed to assess the performance of the proposed MIMO FMCW radar system in estimating respiratory and cardiac metrics under controlled conditions.

\begin{figure}[!th]
\centering
\includegraphics[width=0.89\textwidth]{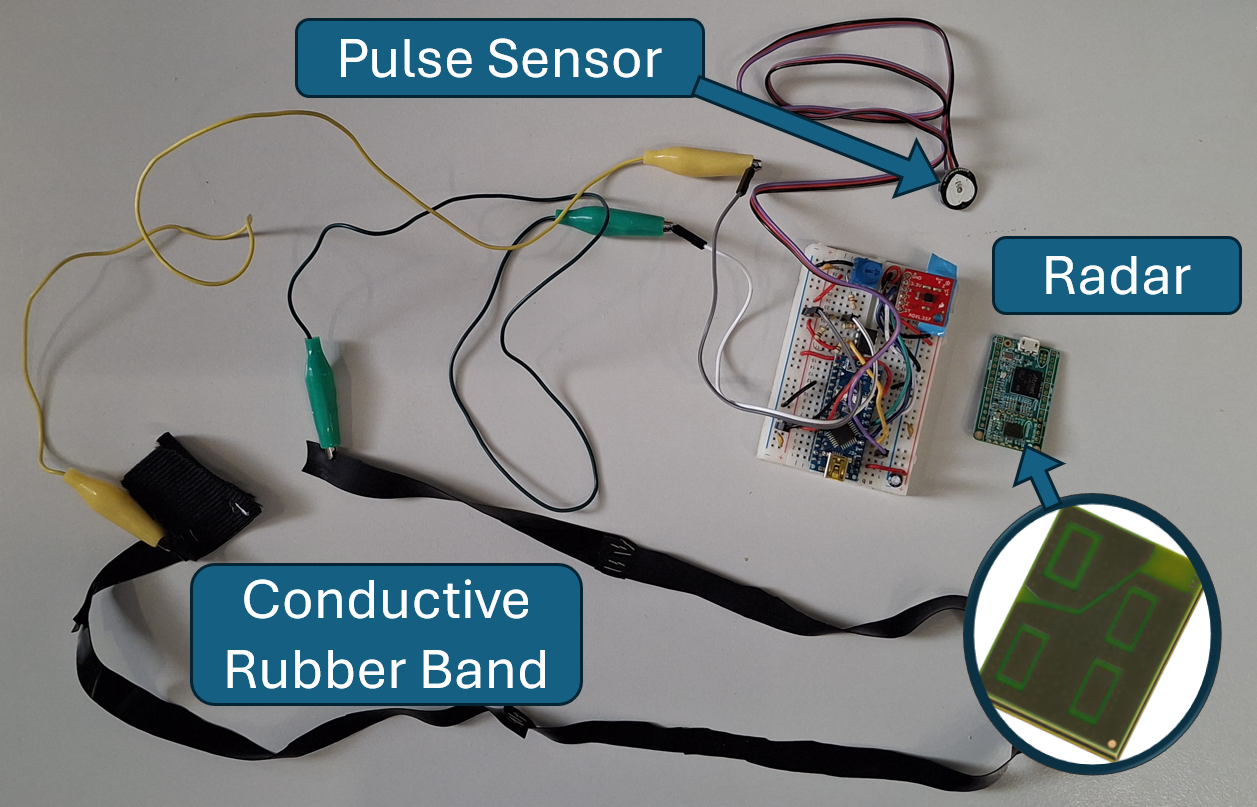}
\caption{Experimental setup showing the sensors used. An Arduino Nano is tasked with collecting the sensors outputs for ground truth. A python script synchronizes the captures from both radar and the Arduino.} \label{fig:experimental_setup_labeled}
\end{figure}

\begin{figure}[!th]
   \begin{minipage}{0.54\textwidth}
     \centering
     \includegraphics[width=.99\linewidth]{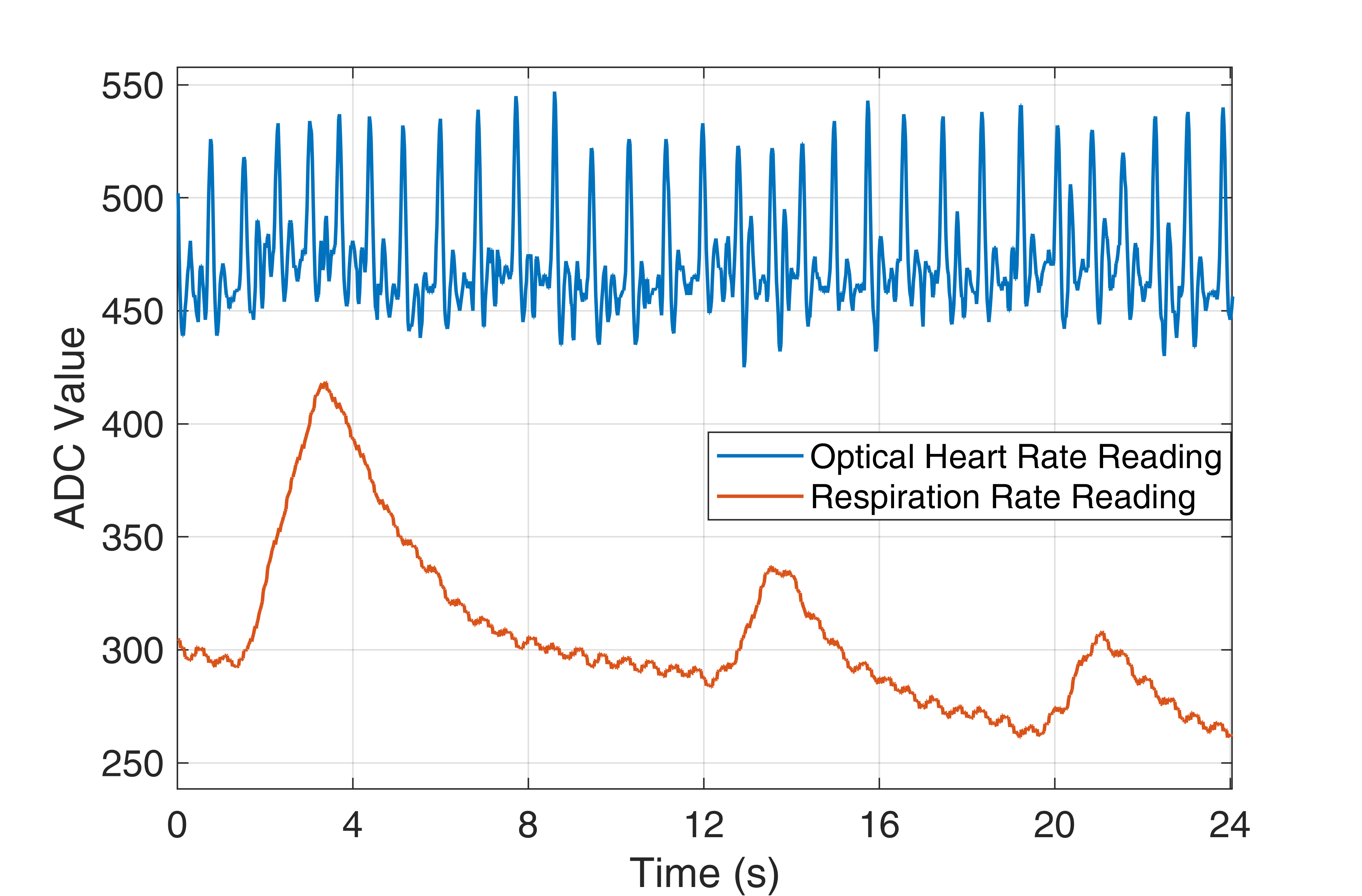}\label{Fig:GT}
   \end{minipage}\hfill
   \begin{minipage}{0.46\textwidth}
     \centering
     \includegraphics[width=.99\linewidth]{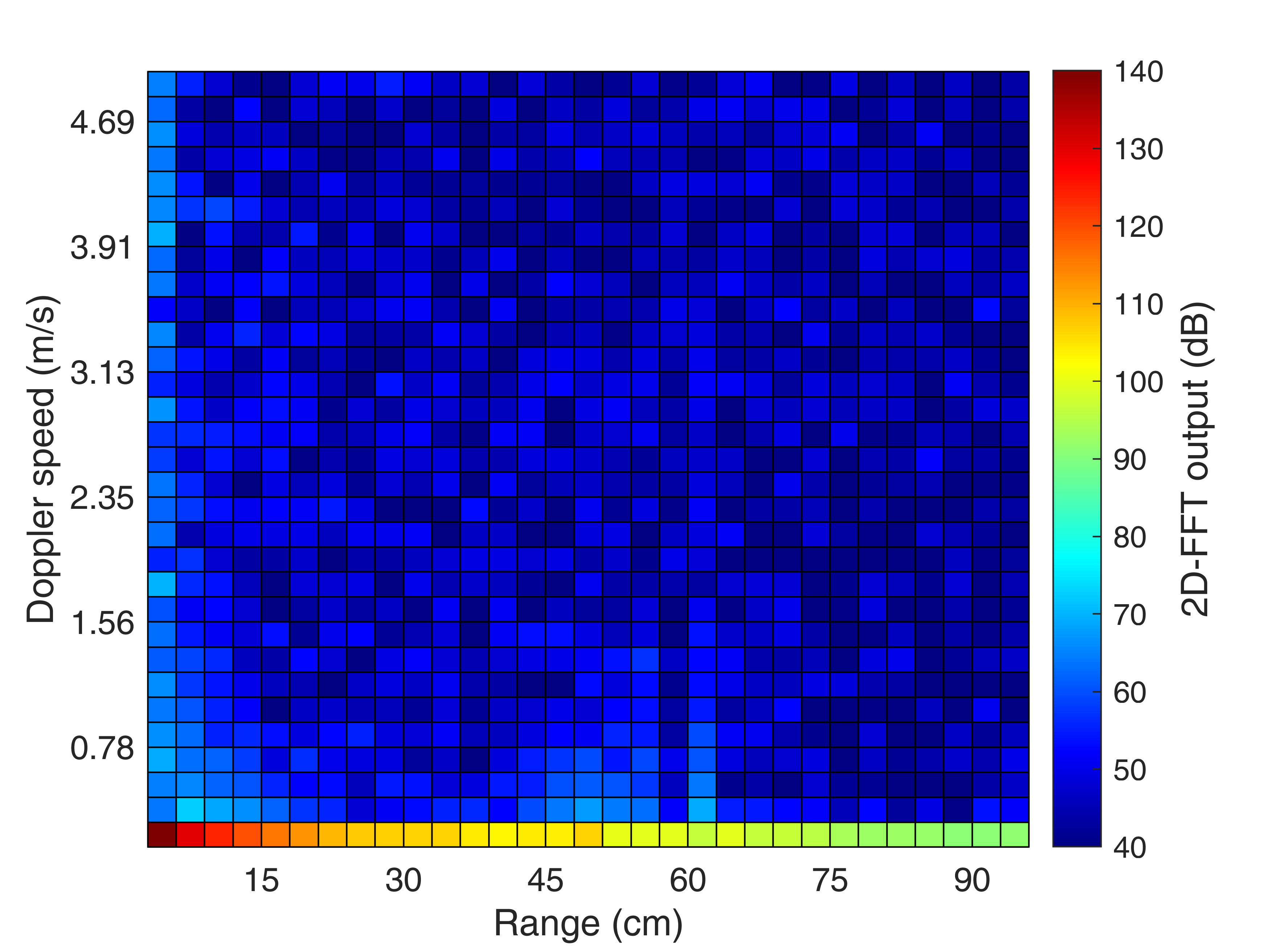}\label{Fig:2DFFT}
   \end{minipage}
   \caption{Ground truth respiration and heart rate readings obtained from the Arduino (left), 2D FFT plot extracted from a single frame of the capture with the person placed at $50\:cm$ from the radar (right).}\label{fig:GT_and_FFT}
\end{figure}

\subsection{Experimental Setup}

\subsubsection{Radar Setup:}
The experimental setup used for the results presented in this paper includes a wide-band mmWave radar sensor as well as two dedicated sensors for respiration rate and heart rate, used for the collection of the ground truth. The mmWave radar used is a BGT60TR13C from Infineon Technologies, which is a small size radar for which an evaluation platform is available~\cite{BGT60TR13C}. The radar IC implements a FMCW Radar with a bandwidth of 5GHz between the frequencies of 58 and 63 GHz. The radar has one TX antenna and three RX antennas arranged in an L configuration to enable 3D DoA estimation, as can be seen from Fig.~\ref{fig:experimental_setup_labeled}. The radar was configured with a chirp repetition time of $124\:\mu s$, number of chirp pulse repetitions of 128 and number of ADC samples per pulse of 128. The chirp waveform was configured to sweep over $5\:GHz$ bandwidth from a starting frequency of $58\:GHz$ to a final frequency of $63\:GHz$. Intermediate frequency gain was set to $28\:dB$ and ADC sample rate was set to 3 MSPS. This resulted in a range resolution of $3\:cm$ and a Doppler resolution of $0.1565\:m/s$. The measurements were conducted on an adult male subject seated comfortably on an office chair and instructed to remain still throughout the acquisition. Data were collected at distances ranging from $30~\mathrm{cm}$ to $150~\mathrm{cm}$, in increments of $10~\mathrm{cm}$, with each recording lasting 2 minutes. These settings were chosen as they were the maximum supported by the sensor while fixing the frame rate at 30 frames per second as well as enable readings up to 2 meters. To evaluate the effect of a higher number of chirps, resulting in improved Doppler resolution, the measures performed ar 30, 50, 70 and 90 cm were repeated with the number of chirps repetitions increased to 256 while samples per chirp were reduced to 64, all other settings were kept the same. This improved the Doppler resolution to $7.8\:cm/s$.

\subsubsection{Ground truth sensors:}
The RR sensor was constructed from a sheet of conductive silicone rubber with resistivity of $120 \:\Omega/cm$, which was cut into strips of $1\:cm$ connected together so as to surround the person. The ends of this rubber are connected to a differential amplifier which reduces DC bias and amplifies the signal to the dynamic range of the Arduino's ADC. 
The HR ground truth was obtained via an optical pulse sensor based on the APDS-9008 optical sensor. It uses a green LED as a source of light and measures how much light is reflected back which correlates to blood concentration which are driven by the heart.

These analog sensors were digitized using an Arduino Nano with a sample frequency of 50 Hz, a python script was used to automate and synchronize the capture from both Arduino and Radar. The first 24 seconds of a ground truth reading alongside an example of a 2D-FFT frame derived from a single antenna of the Radar are given in Fig.~\ref{fig:GT_and_FFT}.

\subsection{Numerical Results}
\begin{figure}[!t]
\centering
\includegraphics[width=0.85\textwidth]{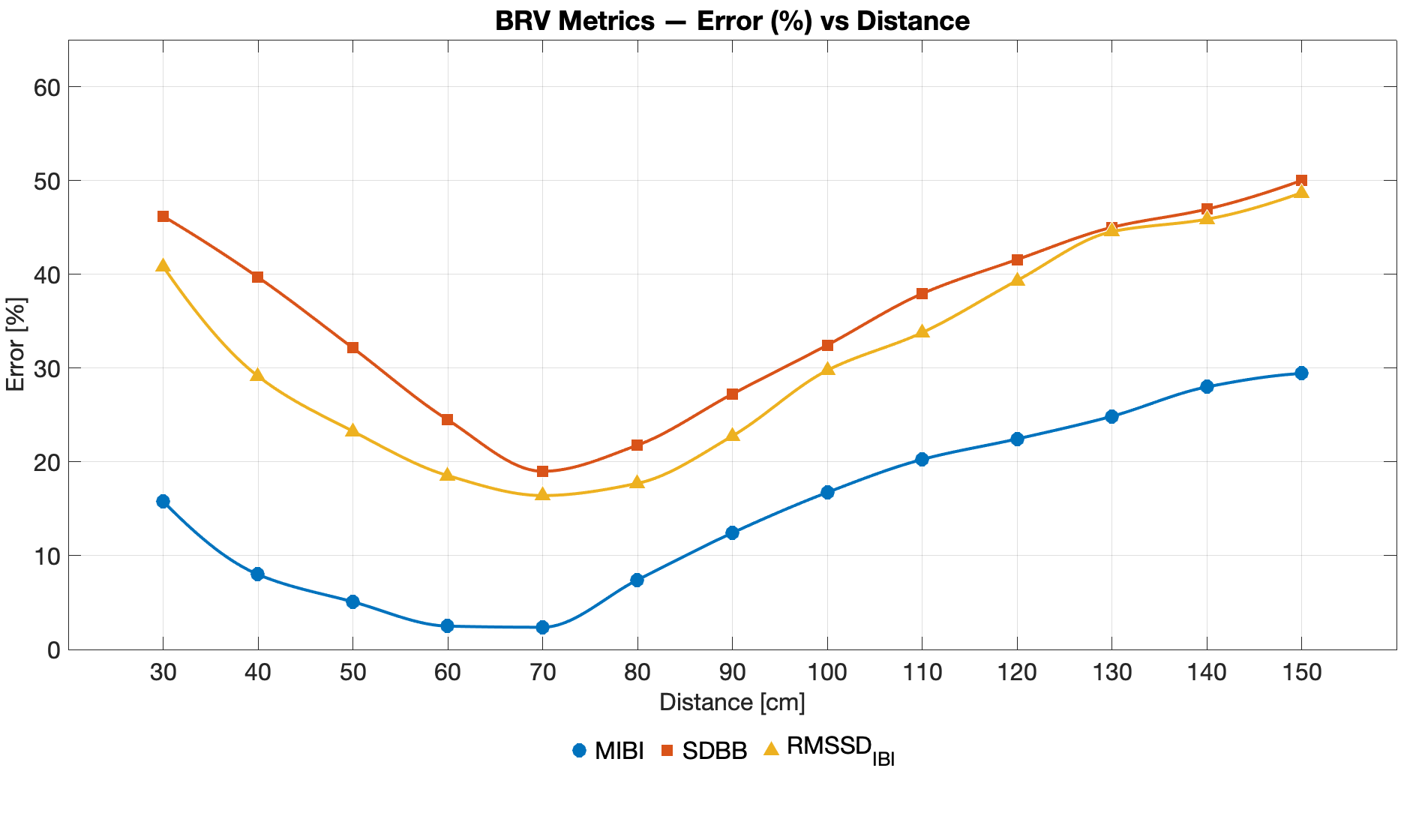}
\caption{RR estimation error versus distance of subject from the radar.} \label{fig:BRV_error_distance}
\end{figure}
\begin{figure}[!t]
\centering
\includegraphics[width=0.85\textwidth]{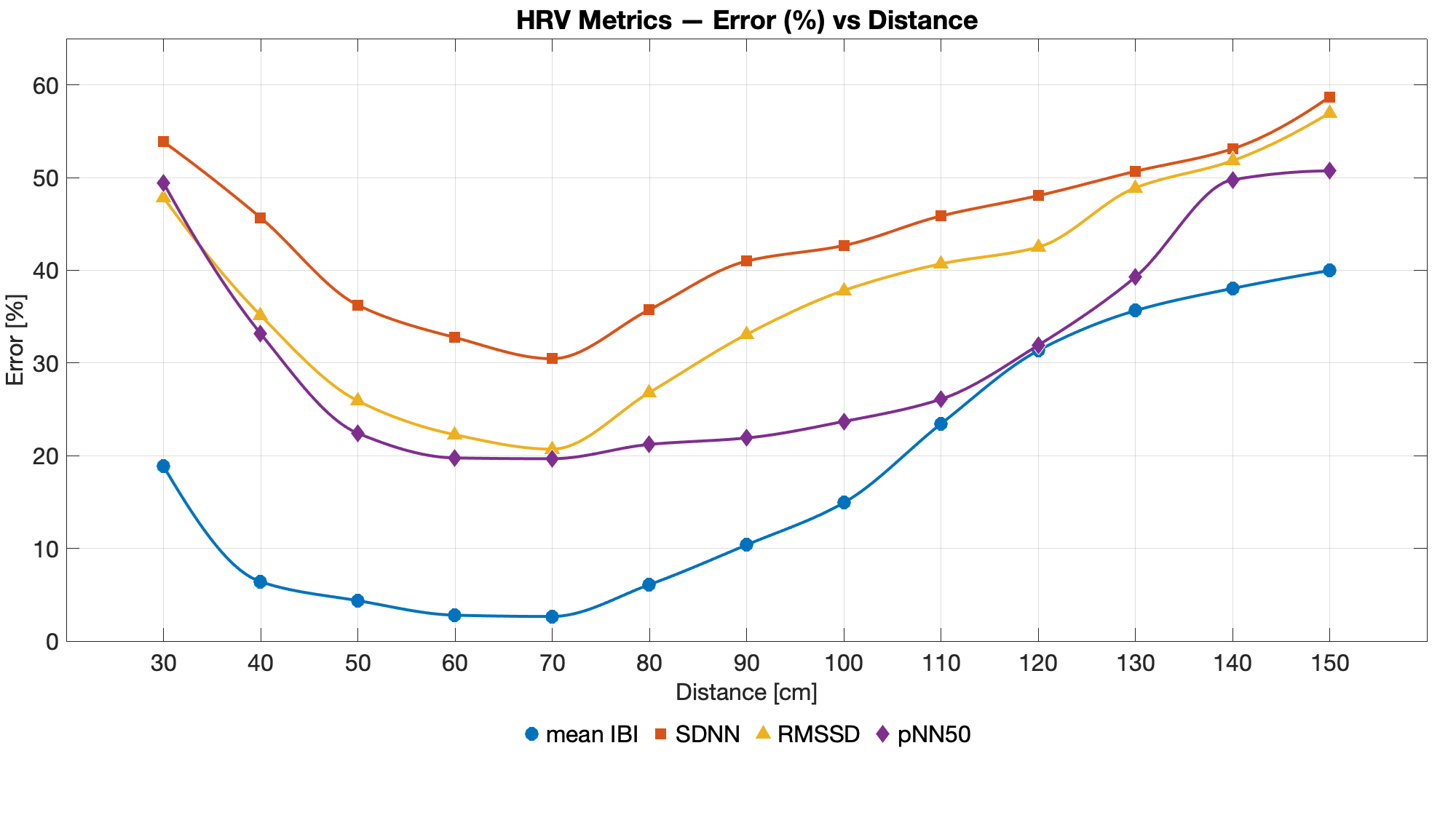}
\caption{HR estimation error versus distance of subject from the radar.} \label{fig:HRV_error_distance}
\end{figure}

The radar performance was evaluated for RR, HR, and their variability metrics (BRV and HRV). Errors are reported as mean absolute error (MAE, mean $\pm$ standard deviation) relative to ground truth from strain sensors (RR) and optical pulse sensors (HR). 

Figures~\ref{fig:BRV_error_distance} and \ref{fig:HRV_error_distance} show the dependence of estimation accuracy on sensing distance. Both RR and HR exhibit a characteristic $U$-shaped error profile, with optimal performance at 70~cm. At this distance, RR estimation achieved a MAE of $0.8 \pm 0.3~bpm$ ( $\approx$ 4.5\% relative error), while HR MAE was $3.2 \pm 1.1~bpm$ ($\approx 4\%$). Respiratory variability metrics at $70~cm$ achieve a MAE of 2\% error for MIBI, 17\% for the standard deviation of breath-to-breath intervals (SDBB), and 20\% for RMSSD. Heart rate variability metrics exhibited higher errors, with SDNN at 30\%, RMSSD 20\%, and pNN50 20\%.  
Performance degrades at shorter distances. At $30~cm$, RR MAE increased to $2.1 \pm 0.7~bpm$ and HR MAE to $7.8 \pm 2.3~bpm$, with variability errors rising to 25--42\%. Degradation at close range is attributed to multipath reflections, near-field coupling, and uneven beam illumination. Beyond $70~cm$, errors increase gradually due to reduced SNR, reaching $1.9 \pm 0.5~bpm$ for RR and $6.4 \pm 1.9~bpm$ for HR at $150~cm$. These results reflect the larger displacement of respiration ($10$--$50~mm$) relative to cardiac motion ($1$--$9~mm$), explaining the higher robustness of RR estimation.  

\begin{figure}[!t]
\centering
\includegraphics[width=0.82\textwidth]{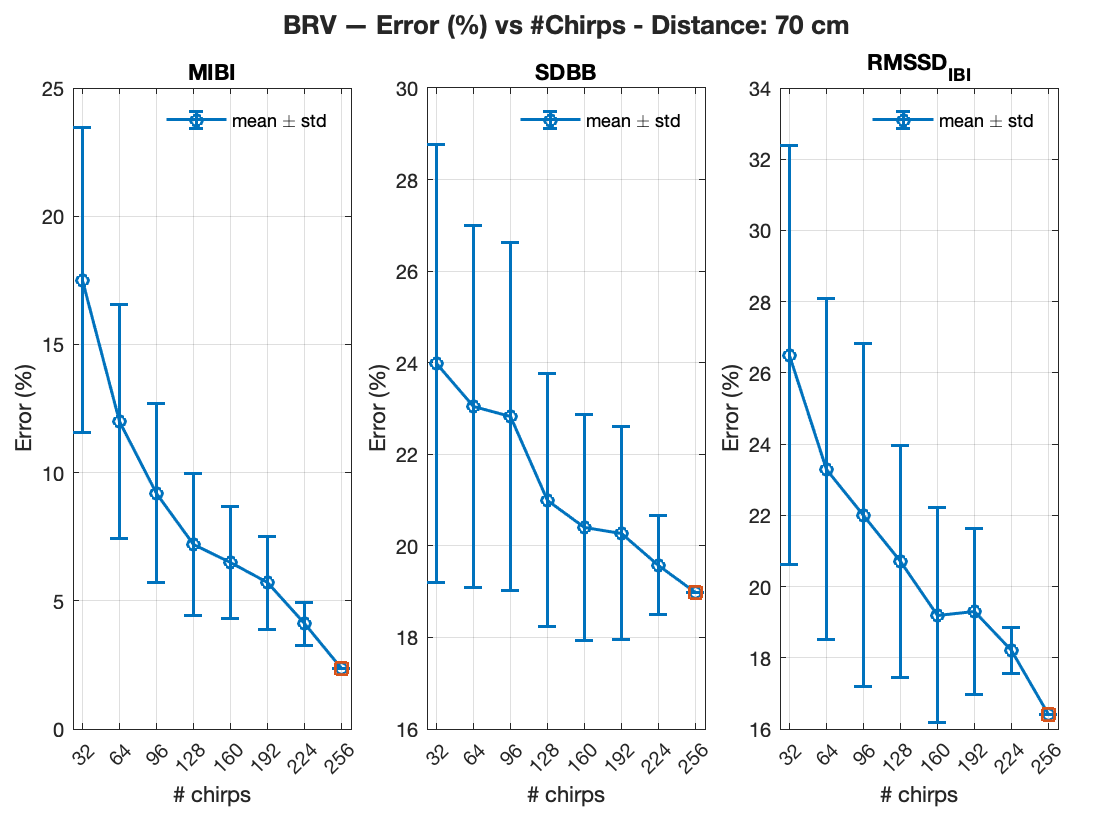}
\caption{RR estimation error versus number of chirps used for subject at a distance of 70cm.} \label{fig:BRV_70cm_chirps}
\end{figure}

\begin{figure}[!th]
\centering
\includegraphics[width=0.80\textwidth]{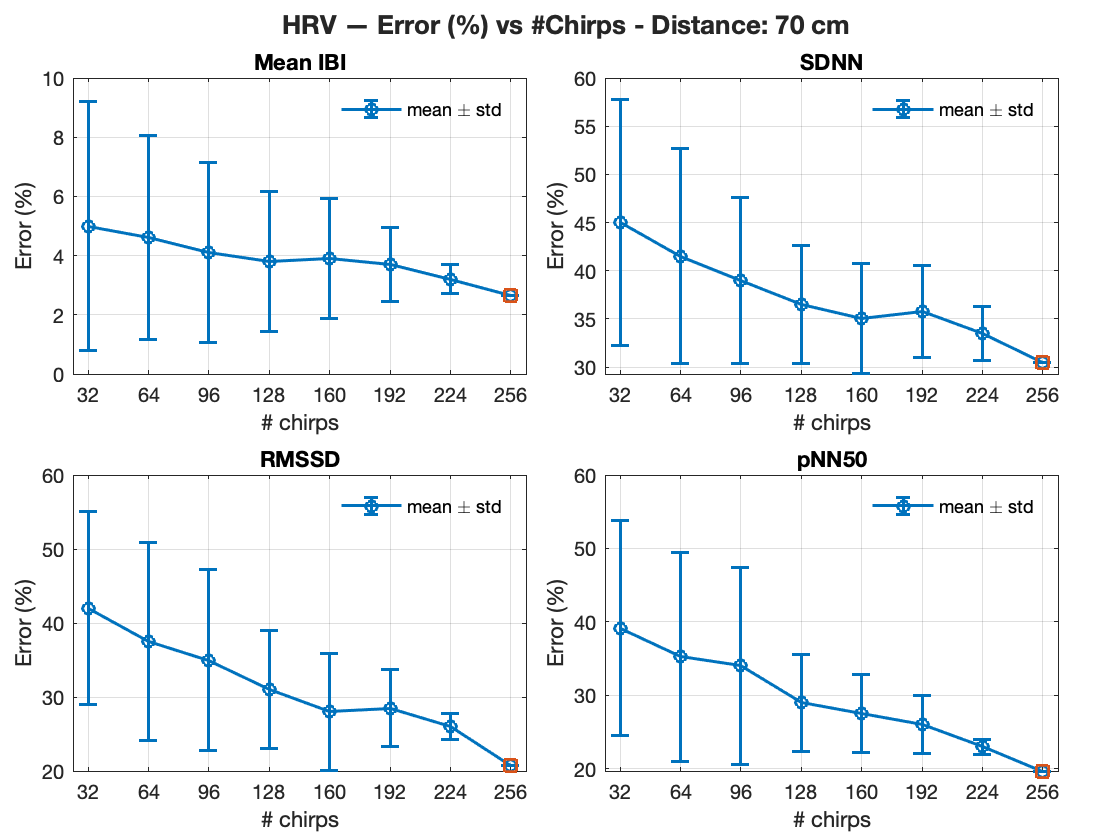}
\caption{HR estimation error versus number of chirps used for subject at a distance of 70cm.} \label{fig:HRV_70cm_chirps}
\end{figure}
The impact of the number of transmitted chirps was evaluated at $70~cm$ (Figures~\ref{fig:BRV_70cm_chirps} and \ref{fig:HRV_70cm_chirps}). RR estimation improves monotonically with increasing chirp count, from $2.8 \pm 0.9~bpm$ at 32 chirps to $0.9 \pm 0.3~bpm$ at 256 chirps, with MIBI error decreasing from 3.2\% to 2.0\%. SDBB and RMSSD remain above 15--20\% even at 256 chirps, indicating limited improvement in variability estimation. 

HR estimation shows a pronounced threshold effect: below 96 chirps, measurements often fail, with MAE exceeding $15~bpm$. Reliable HR estimation requires at least 96 chirps, reducing MAE below $5~bpm$, and at 256 chirps, HR MAE stabilizes near $2.8 \pm 0.9~bpm$. HRV metrics also improve modestly with chirp count, with SDNN decreasing from 52\% at 32 chirps to $28\%$ at 256, RMSSD from $45\%$ to $18\%$, and pNN50 from $48\%$ to $19\%$.  

These results highlight the operational trade-offs inherent to the radar system. Respiratory rate can be measured reliably with MAE below $1~bpm$ over a broad range of distances ($50–120~cm$) when using 96–256 chirps, demonstrating robustness to both positioning and chirp count. In contrast, accurate heart rate estimation is more sensitive, requiring optimal positioning near $70~cm$ and at least 96 chirps to achieve MAE below $5~bpm$. Variability metrics for both respiration and heart remain less precise, with errors typically above 15–30\%, reflecting the challenge of capturing rapid, beat-to-beat or breath-to-breath fluctuations. These findings illustrate a trade-off: increasing chirp count and precise positioning enhance average rate estimation but provide limited gains for instantaneous variability metrics. The system excels at coarse, robust monitoring of RR and HR but is constrained in high-resolution autonomic analysis.

\section{Conclusion}

This study presents a systematic evaluation of a low-cost FMCW MIMO radar for non-contact vital sign monitoring, focusing on respiratory rate (RR) and heart rate (HR) estimation. Extensive experiments quantify how sensing distance and chirp count affect accuracy. RR can be reliably estimated across $50-120~cm$ and chirp counts $\geq96$, with mean absolute errors below $1~bpm$. HR estimation shows stronger dependence on both parameters, requiring optimal positioning near $70~cm$ and at least 96 chirps for stable detection. Variability metrics, including HRV and RRV, remain less accurate (errors $\ge15-30\%$), reflecting challenges in capturing instantaneous physiological fluctuations. Overall, the results reveal an inherent trade-off: the radar enables robust average vital sign estimation but limited beat-to-beat and breath-to-breath resolution. Future works will focus on expansion of the dataset with more subjects and different configurations, including subject movement, and relative algorithm adjustments to overcome these new scenarios.
%Optimized system configuration, combined with multi-domain data fusion and temporal super-resolution, may be essential for achieving fine-grained, clinical-grade monitoring.
\bibliography{ref.bib}
\bibliographystyle{IEEEtran}
%\bibliography{ref}
%\bibliographystyle{splncs04}
%splncs04




% LNCS Homepage, \url{http://www.springer.com/lncs}. Last accessed 4
% \end{thebibliography}

\end{document}